\pdfoutput=1
\documentclass[11pt]{article}

\usepackage[preprint]{acl}

\usepackage{times}
\usepackage{latexsym}

\usepackage[T1]{fontenc}
\usepackage[utf8]{inputenc}

\usepackage{microtype}

\usepackage{inconsolata}

\usepackage{graphicx}

\usepackage[inline]{enumitem}
\title{Mitigating Cyber Risk in the Age of Open-Weight LLMs:\\Policy Gaps and Technical Realities}
\author{Alfonso De Gregorio\\
  Pwnshow\\
  \texttt{adg@pwnshow.com}\\}
  
\begin{document}
\maketitle
\begin{abstract}
Open-weight general-purpose AI (GPAI) models offer significant benefits but also introduce substantial cybersecurity risks, as demonstrated by the offensive capabilities of models like DeepSeek-R1 in evaluations such as MITRE's OCCULT. These publicly available models empower a wider range of actors to automate and scale cyberattacks, challenging traditional defence paradigms and regulatory approaches. This paper analyzes the specific threats—including accelerated malware development and enhanced social engineering—magnified by open-weight AI release. We critically assess current regulations, notably the EU AI Act and the GPAI Code of Practice, identifying significant gaps stemming from the loss of control inherent in open distribution, which renders many standard security mitigations ineffective. We propose a path forward focusing on evaluating and controlling specific high-risk capabilities rather than entire models, advocating for pragmatic policy interpretations for open-weight systems, promoting defensive AI innovation, and fostering international collaboration on standards and cyber threat intelligence (CTI) sharing to ensure security without unduly stifling open technological progress.
\end{abstract}

\section{Introduction}
\subsection{Motivation and Context}
The rapid advancement of general-purpose artificial intelligence (GPAI) models, particularly large language models (LLMs) and multimodal systems, presents both great opportunities and significant challenges. A prominent aspect of this evolution is the proliferation of "open-weight" models – systems whose underlying parameters are publicly released, allowing developers worldwide to download, modify, and deploy them. This open approach fosters transparency, competition, and a "Cambrian explosion" of innovation, enabling researchers and smaller entities to build upon state-of-the-art technology without prohibitive initial investments \cite{brooks:2025}.

However, this openness simultaneously fuels concerns about potential misuse, particularly in the realm of cybersecurity. Evidence suggests that advanced AI models possess significant offensive cyber capabilities. MITRE's OCCULT framework evaluations showed models like DeepSeek-R1 achieving over 90\% accuracy on challenging offensive cyber knowledge tests \cite{Kouremetis:2025}. This capability amplification threatens to lower the barrier for sophisticated cyberattacks – automating phishing campaigns, accelerating malware development, and potentially enabling faster discovery and exploitation of vulnerabilities \cite{Rodriguez:2025}. The release of DeepSeek's R1 model, a capable, efficient, and open-weight reasoning model developed in China, catalyzed these concerns among policymakers, highlighting the potential for unexpected capabilities and the complex geopolitical dimensions \cite{brooks:2025}. Also, specialized models (e.g., \emph{Xanthorox AI}) have already being released in darknet forums in early 2025; they are purposely built from for Offensive Cyber Operations (OCO), and to run on private infrastructure \cite{montalbano:2025}. 

This paper examines the cybersecurity implications of open-weight AI models, analyzing the specific risks they introduce and evaluating the adequacy of current and proposed regulatory frameworks. It argues that while open innovation is crucial, the unique characteristics of open-weight models necessitate a nuanced approach to risk mitigation and regulation, distinct from that applied to closed, API-gated systems.

\subsection{Research Questions and Objectives}
This paper seeks to address the following questions:
\begin{itemize}
\item What specific offensive cyber capabilities are enhanced or enabled by the availability of open-weight GPAI models?

\item How do current regulatory frameworks, such as the EU AI Act and associated Codes of Practice, address the risks posed by publicly available model weights?

\item Why are traditional mitigation strategies often ineffective once model weights are released openly?

\item What policy and technical interventions can effectively mitigate the cybersecurity risks of open-weight AI while preserving the benefits of open innovation?
\end{itemize}

The objective is to provide an evidence-based analysis of the evolving threat landscape, identify critical policy gaps, and offer recommendations for a balanced regulatory approach.

\subsection{Organization}
This paper proceeds as follows: Section 2 provides background on offensive cyber operations, GPAI models (distinguishing open-weight from open-source), and existing research on AI model security. Section 3 details the specific cyber threats amplified by open-weight AI, drawing on recent evaluations like OCCULT. Section 4 analyzes current regulatory frameworks, particularly the EU AI Act \cite{EU:2024} and the GPAI Code of Practice \cite{EC:2025}, highlighting their limitations concerning open-weight models. Section 5 discusses potential defensive innovations and the role of cyber norms. Section 6 presents policy and technical recommendations for mitigating risks while supporting innovation. Section 7 concludes with key takeaways and future research directions.

\section{Background and Literature Review}
\subsection{Offensive Cyber Operations}
Historically, sophisticated offensive cyber operations were the domain of well-resourced state actors and large corporations possessing the necessary expertise, capabilities, and time. However, the cyber threat landscape has become increasingly accessible to a wider range of threat actors. The rise of cybercrime-as-a-service models, including readily available phishing kits and ransomware platforms, has significantly lowered the barrier to entry for less sophisticated actors, including criminal syndicates and hacktivist groups. AI now threatens to further accelerate this trend.

\subsection{General-Puropose AI and Open-Weights Models}
GPAI models, particularly LLMs, are complex systems trained on vast datasets, capable of understanding and generating human-like text, code, and other content. Their development involves significant investment in data, compute, and algorithmic refinement \cite{RAND:2024}. The "weights" of a model represent the learned parameters that encode its capabilities.

A critical distinction exists within the "open" AI ecosystem:\\\\
\emph{Open-Weight Models}: Models where the numerical weights (parameters) are publicly released, allowing users to run, modify, and fine-tune the model locally or on their own infrastructure. Examples include Meta's Llama series \cite{DBLP:journals/corr/abs-2302-13971}, Mistral models \cite{jiang2023mistral7b}, and DeepSeek \cite{deepseekai2025deepseekr1incentivizingreasoningcapability}.\\\\
\emph{Open-Source AI (OSI Definition)}: A stricter definition requiring the release of not only weights but also the training data and code. Most current "open-weight" models do not meet this definition due to proprietary datasets or training methods \cite{OSI:2024}.\\

The access of open-weight models has been driving innovation, transparency, and competition. However, it also complicates systemic risk mitigation, as --- as it will be discussed in Section 4, while addressing relevant policy gaps --- control over the model is lost upon release. Therefore the analysis offered in this paper extends to both open-weight models, and those who are fully open source.

\subsection{Offensive AI in Cybersecurity Literature}
The potential for AI to enhance offensive cyber operations has been a subject of growing research. Studies have explored AI's use in automating vulnerability discovery \cite{Rodriguez:2025}, generating malicious code, crafting sophisticated phishing attacks, and potentially bypassing defenses. Kouremetis et al. \cite{Kouremetis:2025} provide concrete evidence of high LLM proficiency in offensive cyber knowledge through the OCCULT framework. This research underscores the potential dual-use nature of AI and its potential to significantly shift the existing offence-defence asymmetry, where defenders traditionally struggle to keep pace with ever evolving attack vectors. Securing the model weights themselves has been identified as a critical challenge, particularly given their size and the potential for theft or unauthorized copying \cite{RAND:2024}. The reader is referred to the same publication for an introduction to specific vectors by which weights could leak and best practices for partial or minimal release.

\section{Threat Landscape Redefined by Open-Weight AI}
\subsection{Offensive Cyber Potential of Advanced LLMs}
The widespread availability of capable open-weight AI models fundamentally alters the cyber threat landscape by lowering barriers related to expertise, resources, and time for attackers. The offensive potential is vast, extending from the automation and personalization of social engineering, to enhanced vulnerability discovery and exploitation, to scaling misinformation and disinformation campaigns. More specifically, key offensive potentials include:\\\\
\emph{Automated and Personalized Social Engineering}: LLMs excel at generating human-like text, enabling the creation of highly convincing phishing emails, fake profiles, and other social engineering lures at scale, including micro-targeted phishing or deepfake voice calls – a scope far wider than conventional email-based phishing.AI can personalize these attacks based on publicly available or breached data, significantly increasing their effectiveness.\\\\
\emph{Accelerated Malware and Tool Development}: AI can assist attackers in writing, debugging, and modifying malicious code, including ransomware, spyware, and intrusion tools. This reduces the technical skill required and speeds up the development cycle \cite{Kouremetis:2025}. Models can translate functional requirements into code or adapt existing malware to evade specific defenses.\\\\
\emph{Enhanced Vulnerability Discovery and Exploitation}: AI can analyze codebases and systems to identify potential vulnerabilities more rapidly than manual methods. While fully automated zero-day discovery remains challenging, AI significantly aids researchers (and attackers) in fuzzing, symbolic execution, and identifying patterns indicative of flaws.\\\\
\emph{Evasion and Obfuscation}: AI can generate polymorphic malware variants designed to evade signature-based detection or learn techniques to bypass isolation and compartmentalization technologies and behavioral monitoring systems \cite{Rodriguez:2025}.\\\\
\emph{Misinformation and Disinformation at Scale}: While often considered a separate issue, the ability of LLMs to generate plausible but false narratives can be weaponized in cyber campaigns, for instance, to amplify the impact of a data breach or discredit a target organization.

\subsection{MITRE's OCCULT and DeepSeek-R1}
The OCCULT framework provides a methodology for evaluating LLMs on offensive cyber tasks \cite{Kouremetis:2025}. Preliminary results using this framework demonstrated that DeepSeek-R1, an open-weight model, achieved over 90\% accuracy on the TACTL-183 benchmark, covering a wide range of cyber knowledge areas relevant to offense. While multiple-choice tests have limitations, these results indicate a high level of encoded knowledge that could be readily applied by users with offensive intent. The DeepSeek case itself, involving an efficient, capable, open-weight model developed in China, brought these risks into sharp focus for policymakers.

\subsection{Escalation of Capabilities and Actor Profiles}
Open-weight models democratize access to capabilities previously confined to highly skilled actors. This empowers a wide range of actors with greater attack capabilities and potential for misuse. In particular:\\\\
\emph{Less-Skilled Individuals}: Individuals with basic technical knowledge can leverage AI to perform attacks that would otherwise be beyond their reach.\\\\
\emph{Criminal Syndicates}: Groups can increase the scale, speed, and sophistication of their operations (e.g., ransomware, Business Email Compromise (BEC) scams) with reduced overhead.\\\\
\emph{State-Sponsored Actors}: While already possessing significant capabilities, states can use AI to augment their human operators, increase operational tempo, and potentially develop novel attack vectors.\\\\
Finally there exist the \emph{Modification Risks}. In effect the open nature allows actors to fine-tune models with relative ease and resources considerably more limited than those required for training the same model \cite{volkov2024badllama3removingsafety, labonne2024abliteration}, so as to relax or "remove" standard safety or alignment guardrails or specialize them for malicious tasks, as demonstrated by the ease of creating phishing-focused models from general-purpose ones \cite{RAND:2024}. This is a risk the present paper elaborates further in Section 4.2.

\section{Policy Blind Spots in Current Regulatory Frameworks}

\subsection{Overview of Existing Legislation: The EU AI Act \& GPAI Code of Practice}
The EU AI Act represents one of the most comprehensive attempts to regulate AI globally. It is worth mentioning its extraterritorial effects, as the applicable jurisdiction is deliberately inclusive, because the act is triggered whenever any given model is made available in the EU (e.g., published on \emph{Hugging Face}). As such, the EU AI Act has global effects and relevance. It employs a risk-based approach, imposing stricter obligations on "high-risk" AI systems (Article 6 and Annex III). General-purpose AI models are subject to specific transparency requirements (Article 52). Models deemed to pose "systemic risk" (GPAISR) – defined based on capabilities or training compute thresholds – face additional obligations, including model evaluation, risk assessment, mitigation, and cybersecurity protections (Article 51, 55, 56) \cite{EU:2024}.

The "Third Draft of the General-Purpose AI Code of Practice" provides a voluntary, yet detailed, pathway for GPAISR providers to demonstrate compliance with certain AI Act obligations, particularly regarding safety and security \cite{EC:2025} It outlines commitments for systemic risk assessment (Commitment II.2-II.5), technical mitigations (II.6-II.7), governance (II.8-II.16), including cybersecurity best practices (Measure II.7.1, referencing RAND SL3), protection against model weight theft (Measure II.7.3), and insider threat programs (Measure II.7.5) \cite{EC:2025}.

\subsection{Systemic Risk vs. Open-Source Culture: The Mitigation Gap}
A fundamental challenge arises when applying regulations (e.g., EU AI Act) – however sensible and well reasoned but – designed primarily for closed, proprietary systems to open-weight models. Once model weights are publicly distributed:
\begin{itemize}
\item \emph{Control is Lost}: The original developer loses direct control over how the model is used, modified, or deployed.

\item \emph{Mitigations Become Ineffective}: Many standard security mitigations become largely irrelevant. Rate-limiting APIs, hardware enclaves (e.g., TEEs), access controls, watermarking, and monitoring of downstream use are bypassed when users can run the model independently. Commitments in the GPAI Code of Practice regarding securing unreleased weights (Commitment II.7) or post-market monitoring (Measure II.4.14) do not address the risks once weights are released.

\item \emph{Modification is Easy}: As shown with fine-tuning, safety features, embedded in the model, and safety alignment, often implemented via fine-tuning or Reinforcement Learning from Human Feedback (RLHF), can be trivially sidestepped or circumvented by adversaries with access to the weights (Section 3.3). Mandating developers prevent modification (an implicit requirement in some liability proposals) is technically infeasible for open-weight released publicly, as models run on premises.

\item \emph{Attribution is Difficult}: Tracing malicious activity back to the use of a specific open-weight model becomes challenging, as appropriate mitigation can be disabled (e.g., watermarking) and activity related to the model deployment is kept off-the-record.
\end{itemize}

\subsection{Case Studies of Policy Gaps}
The EU AI Act's Open Source Exemption: The Act generally exempts models released under "free and open-source licences" from most obligations, unless they are deemed GPAISR (Article 2(10e), Recital 12b). However, the definition of "free and open-source" remains ambiguous. It can be interpreted either strictly or loosely, but both interpretations are problematic in distinct ways. If interpreted strictly according to the OSI definition (requiring data release), most current open-weight models (Llama, Mistral, DeepSeek) would not qualify for the exemption and would face the full GPAISR obligations, potentially stifling their release. If interpreted loosely as "open-weight," then potentially highly capable, risky models might escape necessary scrutiny. The GPAI Code of Practice draft does not explicitly address this exemption nuance.

Furthermore, there are regulatory scenarios characterized by implicit restrictions. Legislation imposing broad liability standards (e.g., "reasonable care" against critical risks, as seen in the Californian SB 1047 proposal \cite{SB1047:2024}) or requiring monitoring/control creates disproportionate burdens for open-weight developers who lack visibility and control over downstream use, implicitly favouring closed models.

Finally there is conflation between obligations relating to the models or to systems. Requirements like content watermarking, when imposed at the model level, are difficult to enforce robustly for open-weight models as the watermarking mechanism can often be removed or bypassed. Such obligations would be more appropriately placed at the application/system level.

Overall, the current regulatory landscape, including the EU AI Act and its draft Code of Practice, while comprehensive for closed systems, exhibits significant blind spots regarding the unique risks and technical realities of publicly released open-weight models. Section 5 will explore what can be done to address the current regulatory and technical shortcomings.

%\section{Cyber Norms and Defensive Innovation}
\section{Defensive Innovation and International Cooperation}
\subsection{Defensive Technologies -- "AI to Fight AI"}
While AI enhances offensive capabilities, it also offers powerful tools for cyber defense. A detail treatment is outside the scope of the present work. However, we present a brief overview of some key innovations in this space:\\\\
\emph{AI-Powered Threat Detection}: Utilizing machine learning for advanced anomaly detection in network traffic, user behavior, and system activity to identify sophisticated intrusions that evade traditional signatures \cite{Rodriguez:2025}.\\
\emph{Automated Incident Response}: AI systems capable of automatically analyzing alerts, orchestrating containment measures, and assisting human analysts in responding to breaches more rapidly.\\\\
\emph{Predictive Threat Intelligence}: AI models analyzing vast datasets to identify emerging threats, predict attacker tactics, techniques, and procedures (TTPs), and proactively harden defenses.\\\\
\emph{Vulnerability Management}: AI assisting in static and dynamic code analysis to identify vulnerabilities before they can be exploited.\\

However, these defensive AI systems face challenges, including adversarial evasion (i.e., attackers designing malware specifically to fool AI detectors), the need for vast amounts of high-quality training data, and the risk of false positives overwhelming incident response teams. Policymakers should promote the development of robust AI enabled defenses by incentivizing the creation of specialized datasets and greater cooperation in this space.

\subsection{International Cooperation and Intelligence Sharing}
We contend that mitigating the risks of powerful, globally accessible AI requires international cooperation, which shall follow multiple avenues:\\\\
\emph{ Cyber Threat Intelligence (CTI) Sharing}: Establishing robust platforms and protocols for rapid sharing of intelligence on AI-driven attacks and vulnerabilities among nations and private sector entities (e.g., through NATO CCDCOE, ENISA, CISA). In a way, access to models capable of OCOs, shall pressure intelligence work to be less self-referential, and to benefit a wider range of constituencies.\\\\
\emph{Developing Common Standards}: Collaborating on international standards for AI safety and security evaluations, particularly for models with offensive potential. Framework such as MITRE's OCCULT offer the means to assess the extent to which an offensive potential exist.\\\\
\emph{Joint Research and Development}: Pooling resources for research into AI safety, security, and defensive AI applications. As an example of possible lines of work, further research would be advisable in: \begin{enumerate*} \item novel architectures whose systemic risk mitigation are less prone to be bypassed when released in open source; \item \emph{semi-open} distribution models, where certain pre-trained layers are locked, or specialized functionalities are gated, or where multiple domain-specific weight sets can be combined by authorized users; \item advanced watermarking approaches.\end{enumerate*}\\\\
\emph{Establishing Norms of Behavior}: Promoting international norms against the development and deployment of AI for malicious cyber operations, although enforcement remains a significant challenge in an inherently global setting.
\section{Policy and Technical Recommendations}
Addressing the cybersecurity risks of open-weight AI requires a multi-faceted approach that avoids stifling innovation unnecessarily. Building on the principle that restrictions should be the exception, targeted, proportionate, and intentional, we propose a range of measures encompassing: controlled releases, incentives set, and balanced approach between innovation and security oversight. The following subsections detail each area.
\subsection{Targeted Gaiting and Controlled Release for High-Risk Capabilities}
\emph{Capability-Specific Controls}: Instead of restricting entire models, focus regulations or release controls on specific, high-risk capabilities demonstrated through rigorous evaluation (e.g., consistent success in automated exploit generation for critical vulnerabilities, as assessed by frameworks like \cite{Kouremetis:2025} or \cite{Rodriguez:2025}. This requires investment in robust evaluation infrastructure (e.g., NIST, AI Safety Institute).\\\\
\emph{Tiered Access Models}: Explore models where core weights are open, but highly sensitive components or fine-tuning capabilities related to offensive cyber operations are subject to stricter access controls or licensing, potentially requiring KYC or adherence to specific use terms. This is technically challenging but warrants investigation.\\\\
\emph{Nuanced Focus on Compute Thresholds}: While compute thresholds (as used in the EU AI Act and GPAI Code of Practice) offer a proxy for capability, they are imperfect. Policy should allow for adjustments based on demonstrated capabilities (or lack thereof) through evaluation, avoiding rigid application that captures inefficient or harmless models, or misses highly efficient dangerous ones. AI Red Teaming complements frameworks for the assessment of capabilities.
\subsection{Incentives for Responsible Practices and Transparency}
\emph{Responsible Labeling}: Encourage or mandate clear labeling by developers regarding known offensive capabilities or limitations identified during pre-release testing. It comes as no surprise that GPAI models are capable of OCO, if specific literature is included in the training set.\\\\
\emph{Partial Obfuscation/Redaction}: Explore techniques for releasing models where specific functionalities known to be high-risk for misuse are partially obfuscated or require additional steps/keys to unlock, increasing friction for malicious actors without fully closing the model.\\\\
\emph{Safe Harbors and Incentives}: Provide liability safe harbors or other incentives (e.g., tax credits, grants) for developers who adhere to best practices in safety evaluation, transparency (e.g., detailed model cards, dataset summaries as per GPAI Code of Practice), and responsible release protocols, particularly for open-weight models.
\subsection{Bridging Open Innovation and Security Oversight}
Bridging innovation and security oversight is possible if, in the overall balanced risk mitigation approach, the following actions are considered too:\\\\
\emph{Clarify "Open Source" Definition for Regulatory Purposes}: Policymakers, especially the EU AI Office interpreting the AI Act, should adopt a pragmatic definition for exemptions that recognizes the value of open-weight releases (even without full data release), ensuring developers like Meta or Mistral are not inappropriately burdened with obligations designed for closed giants.\\\\
\emph{Prioritize Downstream Regulation}: Focus primary regulatory efforts on the deployment and application of AI systems in specific contexts (finance, healthcare, critical infrastructure) rather than imposing excessive burdens on foundational model developers, especially open ones. Address harmful outputs and uses through sector-specific regulations or targeted rules on high-risk applications.\\\\
\emph{Strengthen Public Evaluation Capacity}: Invest heavily in public institutions (NIST, AI Safety Institute) to develop standardized benchmarks (building on work like Kouremetis et al., Rodriguez et al.), conduct independent evaluations, monitor the landscape, and provide guidance, enabling responsive and evidence-based regulation.\\\\
\emph{Collaborative Red-Teaming}: Foster public-private partnerships for rigorous, pre-release red-teaming of significant open-weight models focused specifically on cyber misuse potential.

\section{Conclusions and Further Work}
\subsection{Key Takeaways}
Open-weight GPAI models represent a paradigm shift, accelerating innovation but also significantly altering the cybersecurity threat landscape. They demonstrably possess, or are rapidly acquiring, capabilities that can automate and scale offensive cyber operations, lowering barriers for a wider range of threat actors. Current regulatory frameworks, including the EU AI Act and its draft GPAI Code of Practice, primarily designed for closed systems, exhibit critical blind spots concerning the unique challenges posed by models whose weights are publicly distributed. Traditional security mitigations often become ineffective once control over the model artifact is lost. DeepSeek-R1 serves as a potent example of the capabilities emerging from unexpected sources, demanding a proactive and nuanced policy response.

\subsection{Future Research Directions}
Continued research is crucial to navigate this evolving landscape:
\begin{itemize}
\item Robust Evaluation Frameworks: Refining benchmarks like OCCULT and Rodriguez et al. to better measure real-world offensive potential across diverse cyber tasks and TTPs.

\item Technical Mitigation for Open Models: Investigating novel techniques for embedding safeguards robust against removal/bypass in open-weight settings, verifiable watermarking, or capability limitations within open-weight models.

\item Economic Impact Analysis: Quantifying the potential cost reduction AI offers attackers across different cyberattack chains to better prioritize defenses.

\item Understanding Fine-Tuning Risks: Deeper investigation into the ease and effectiveness of fine-tuning open-weight models to bypass safety alignments or specialize them for malicious purposes.

\item Interdisciplinary Collaboration: Fostering closer ties between AI researchers, cybersecurity experts, legal scholars, and policymakers to develop effective governance solutions.
\end{itemize}

\subsection{Closing Remarks}
The proliferation of capable open-weight AI models presents a complex challenge requiring immediate and coordinated global attention. While the risks, particularly in cybersecurity, are significant, the benefits of open innovation in AI – fostering competition, transparency, and accessibility – are equally important. A balanced approach is necessary, one that avoids reactionary restrictions and instead focuses on targeted interventions, downstream regulation, robust public evaluation capacity, and international cooperation. By harnessing AI for defense and developing adaptive governance frameworks, we can strive to mitigate the risks while realizing the transformative potential of this technology. The journey requires ongoing vigilance, evidence-based policymaking, and a commitment to bridging the gap between rapid technological advancement and effective security oversight.

\section*{Limitations}
Both the technology landscape and regulatory frameworks are evolving rapidly. Therefore, the present evidence-based analysis may lose alignment with the reference environment as it changes. Furthermore, this paper primarily synthesises existing research and publicly available evaluations, such as MITRE's OCCULT framework, to analyse the current threat landscape and policy gaps. While this approach provides a broad overview, the paper does not introduce new primary empirical data or original comparative analyses of offensive capabilities across different models. The validation of specific capability claims therefore relies on the robustness of these external studies. Future work could strengthen these arguments by incorporating original empirical investigations into the actual usage trends and evolving offensive potential of open-weight LLMs.

\section*{Acknowledgements}
The authors would like to express their sincere gratitude to the anonymous reviewers for their insightful comments and constructive feedback on earlier revisions of this manuscript. Their suggestions were invaluable in helping us refine our arguments and improve the overall quality and clarity of this paper.

\bibliography{custom}

\begin{thebibliography}{14}
\providecommand{\natexlab}[1]{#1}

\bibitem[{Brooks(2025)}]{brooks:2025}
Ben Brooks. 2025.
\newblock \href {https://www.youtube.com/watch?v=zdjdeMxg29Y} {Open source
  lawfare: Ai regulation after deepseek [seminar]}.

\bibitem[{{California Senate}(2024)}]{SB1047:2024}
{California Senate}. 2024.
\newblock \href
  {https://leginfo.legislature.ca.gov/faces/billNavClient.xhtml?bill_id=202320240SB1047}
  {Sb-1047 safe and secure innovation for frontier artificial intelligence
  models act.}

\bibitem[{DeepSeek-AI et~al.(2025)DeepSeek-AI, Guo, Yang, Zhang, Song, Zhang,
  Xu, Zhu, Ma, Wang, Bi, Zhang, Yu, Wu, Wu, Gou, Shao, Li, Gao, Liu, Xue, Wang,
  Wu, Feng, Lu, Zhao, Deng, Zhang, Ruan, Dai, Chen, Ji, Li, Lin, Dai, Luo, Hao,
  Chen, Li, Zhang, Bao, Xu, Wang, Ding, Xin, Gao, Qu, Li, Guo, Li, Wang, Chen,
  Yuan, Qiu, Li, Cai, Ni, Liang, Chen, Dong, Hu, Gao, Guan, Huang, Yu, Wang,
  Zhang, Zhao, Wang, Zhang, Xu, Xia, Zhang, Zhang, Tang, Li, Wang, Li, Tian,
  Huang, Zhang, Wang, Chen, Du, Ge, Zhang, Pan, Wang, Chen, Jin, Chen, Lu,
  Zhou, Chen, Ye, Wang, Yu, Zhou, Pan, Li, Zhou, Wu, Ye, Yun, Pei, Sun, Wang,
  Zeng, Zhao, Liu, Liang, Gao, Yu, Zhang, Xiao, An, Liu, Wang, Chen, Nie,
  Cheng, Liu, Xie, Liu, Yang, Li, Su, Lin, Li, Jin, Shen, Chen, Sun, Wang,
  Song, Zhou, Wang, Shan, Li, Wang, Wei, Zhang, Xu, Li, Zhao, Sun, Wang, Yu,
  Zhang, Shi, Xiong, He, Piao, Wang, Tan, Ma, Liu, Guo, Ou, Wang, Gong, Zou,
  He, Xiong, Luo, You, Liu, Zhou, Zhu, Xu, Huang, Li, Zheng, Zhu, Ma, Tang,
  Zha, Yan, Ren, Ren, Sha, Fu, Xu, Xie, Zhang, Hao, Ma, Yan, Wu, Gu, Zhu, Liu,
  Li, Xie, Song, Pan, Huang, Xu, Zhang, and
  Zhang}]{deepseekai2025deepseekr1incentivizingreasoningcapability}
DeepSeek-AI, Daya Guo, Dejian Yang, Haowei Zhang, Junxiao Song, Ruoyu Zhang,
  Runxin Xu, Qihao Zhu, Shirong Ma, Peiyi Wang, Xiao Bi, Xiaokang Zhang,
  Xingkai Yu, Yu~Wu, Z.~F. Wu, Zhibin Gou, Zhihong Shao, Zhuoshu Li, Ziyi Gao,
  and 181 others. 2025.
\newblock \href {https://arxiv.org/abs/2501.12948} {Deepseek-r1: Incentivizing
  reasoning capability in llms via reinforcement learning}.
\newblock \emph{Preprint}, arXiv:2501.12948.

\bibitem[{{European Commission}(2025)}]{EC:2025}
{European Commission}. 2025.
\newblock \href
  {https://digital-strategy.ec.europa.eu/en/library/third-draft-general-purpose-ai-code-practice-published-written-independent-experts}
  {Third draft of the general-purpose ai code of practice: Commitments by
  providers of general-purpose ai models with systemic risk - safety and
  security section. draft document.}

\bibitem[{{European Union}(2024)}]{EU:2024}
{European Union}. 2024.
\newblock \href
  {https://eur-lex.europa.eu/legal-content/EN/TXT/?uri=CELEX:32024R1689}
  {\emph{Regulation (EU) 2024/1689 of the European Parliament and of the
  Council of 13 June 2024 laying down harmonised rules on artificial
  intelligence and amending Regulations (EC) No 300/2008, (EU) No 167/2013,
  (EU) No 168/2013, (EU) 2018/858, (EU) 2018/1139 and (EU) 2019/2144 and
  Directives 2014/90/EU, (EU) 2016/797 and (EU) 2020/1828 (Artificial
  Intelligence Act).}}
\newblock OJ L, 2024/1689.

\bibitem[{Jiang et~al.(2023)Jiang, Sablayrolles, Mensch, Bamford, Chaplot,
  de~las Casas, Bressand, Lengyel, Lample, Saulnier, Lavaud, Lachaux, Stock,
  Scao, Lavril, Wang, Lacroix, and Sayed}]{jiang2023mistral7b}
Albert~Q. Jiang, Alexandre Sablayrolles, Arthur Mensch, Chris Bamford,
  Devendra~Singh Chaplot, Diego de~las Casas, Florian Bressand, Gianna Lengyel,
  Guillaume Lample, Lucile Saulnier, Lélio~Renard Lavaud, Marie-Anne Lachaux,
  Pierre Stock, Teven~Le Scao, Thibaut Lavril, Thomas Wang, Timothée Lacroix,
  and William~El Sayed. 2023.
\newblock \href {https://arxiv.org/abs/2310.06825} {Mistral 7b}.
\newblock \emph{Preprint}, arXiv:2310.06825.

\bibitem[{Kouremetis et~al.(2025)Kouremetis, Dotter, Byrne, Martin, Michalak,
  Russo, Threet, and Zarrella}]{Kouremetis:2025}
M.~Kouremetis, M.~Dotter, A.~Byrne, D.~Martin, E.~Michalak, G.~Russo,
  M.~Threet, and G.~Zarrella. 2025.
\newblock \href {https://arxiv.org/abs/cs.CR/arXiv:2502.15797} {Occult:
  Evaluating large language models for offensive cyber operation capabilities.}

\bibitem[{Labonne(2024)}]{labonne2024abliteration}
Maxime Labonne. 2024.
\newblock \href {https://huggingface.co/blog/mlabonne/abliteration} {Uncensor
  any llm with abliteration}.

\bibitem[{Montalbano(2025)}]{montalbano:2025}
E.~Montalbano. 2025.
\newblock \href
  {https://www.darkreading.com/threat-intelligence/autonomous-genai-attacker-platform-chat}
  {Autonomous, genai-driven attacker platform enters the chat}.
\newblock \emph{Dark Reading}, 12:2025.

\bibitem[{Nevo et~al.(2024)Nevo, Lahav, Karpur, Bar-On, Bradley, and
  Alstott}]{RAND:2024}
S.~Nevo, D.~Lahav, A.~Karpur, Y.~Bar-On, H.~A. Bradley, and J.~Alstott. 2024.
\newblock \href {https://www.rand.org/pubs/research_reports/RRA2849-1.html}
  {\emph{Securing AI Model Weights: Preventing Theft and Misuse of Frontier
  Models}}.
\newblock RAND Corporation. RR-A2849-1.

\bibitem[{OSI(2024)}]{OSI:2024}
OSI. 2024.
\newblock \href {https://opensource.org/ai/open-source-ai-definition} {The open
  source ai definition – 1.0}.

\bibitem[{Rodriguez et~al.(2025)Rodriguez, Popa, Flynn, Liang, Dafoe, and
  Wang}]{Rodriguez:2025}
M.~Rodriguez, R.~A. Popa, F.~Flynn, L.~Liang, A.~Dafoe, and A.~Wang. 2025.
\newblock \href {https://doi.org/10.48550/arXiv.2503.11917} {A framework for
  evaluating emerging cyberattack capabilities of ai}.

\bibitem[{Touvron et~al.(2023)Touvron, Lavril, Izacard, Martinet, Lachaux,
  Lacroix, Rozi{\`{e}}re, Goyal, Hambro, Azhar, Rodriguez, Joulin, Grave, and
  Lample}]{DBLP:journals/corr/abs-2302-13971}
Hugo Touvron, Thibaut Lavril, Gautier Izacard, Xavier Martinet, Marie{-}Anne
  Lachaux, Timoth{\'{e}}e Lacroix, Baptiste Rozi{\`{e}}re, Naman Goyal, Eric
  Hambro, Faisal Azhar, Aur{\'{e}}lien Rodriguez, Armand Joulin, Edouard Grave,
  and Guillaume Lample. 2023.
\newblock \href {https://doi.org/10.48550/ARXIV.2302.13971} {Llama: Open and
  efficient foundation language models}.
\newblock \emph{CoRR}, abs/2302.13971.

\bibitem[{Volkov(2024)}]{volkov2024badllama3removingsafety}
Dmitrii Volkov. 2024.
\newblock \href {https://arxiv.org/abs/2407.01376} {Badllama 3: removing safety
  finetuning from llama 3 in minutes}.
\newblock \emph{Preprint}, arXiv:2407.01376.

\end{thebibliography}

%\appendix
%
%\section{Example Appendix}
%\label{sec:appendix}
%
%This is an appendix.
%
\end{document}